# Evolution and the Calcite Eye Lens

Vernon L. Williams -- vernvlw81@gmail.com

Abstract: Calcite is a uniaxial, birefringent crystal, which in its optically transparent form, has been used for animal eye lenses, the trilobite being one such animal. Because of the calcite birefringence there is a difficulty in using calcite as a lens. When the propagation direction of incoming light is not exactly on the c-axis, the mages blur. In this paper, calcite blurring is evaluated, and the non-blurring by a crystallin eye lens is compared to a calcite one.

# 1. Introduction

Trilobites were so plentiful in ancient seas they were called "the butterflies of the sea". Fortey has said that if you were able to scuba dive during the Ordovician Period some 450,000,000 years ago, you would have found trilobites swarming everywhere [For 2004]. The fossil record confirms that the trilobites were very prolific in the seas. In Figure 1 of Clarkson [Clar 2006] there is a depiction of the various types of trilobites that existed during a 250 million year period. Trilobites probably had the oldest preserved visual system [Clar 2006, Par 2003, pp 221].

     The visual system of the trilobites contained and calcite eye lens. Today there is only one known animal that has a calcite eye lens. This animal is the brittlestar [Alz 2001]. The question arises: why did so many species that evolved after the trilobite not have calcite for an eyes lens?

     A number of researchers have evaluated the optics of the trilobite eye lens [Hor 1989, Hor 1993, Lev 1993]. These papers do contain optical details for the operation of the trilobite eye. However, the papers do not discuss the problems that arise because of calcite. Josef Gal et al [Gal 1999], in this paper "Image formation by bifocal lenses in a trilobite eye" developed a theory and discussed the two focal planes in the trilobite eye, *Dalmanitina socialis*. (Although not discussed in [Gal 1999], it is possible that calcite birefringence helped cause the two focal plane positions.) The *Dalmanitina socialis* trilobite with its at a spherical back surface evolved late in the era of the of trilobite's reign. This was easily seen in the fossil record because the calcite lens did not decay. In his book *Trilobites* [Lev 1993] Levi-Setti discusses this lens that has an aspherical back surface. He postulated that the aspherical surface had the purpose of reducing spherical aberration. The *Dalmanitina socialis* unlike most trilobites had fewer large lens facets but each with larger field-of-view.

# 2. Assumptions

This paper assumes:

- The trilobite eye lens is modeled to be a singlet lens of calcite.

- The material of the eye lens is calcite and with the c-axis parallel to the optical axis.

- The lens conventions and nomenclature used are taken from Nussbaum texts [Nus 1978 and 1889].

- Because calcite is a birefringent, uniaxial crystal, Snell's Law does not hold if light is not impinging directly on the c-axis. The index of refraction for calcite varies with of light propagated angle through the calcite. A calculation will be made later that shows how the index of refraction varies with the propagation angle.

- All incoming light is unpolarized.

# 3. Optical Qualities of Calcite

The eyes of the trilobite have been preserved in the fossil record because the eye lens and surrounding structure were made of calcite, which does not decay. The calcite eye for early trilobites was an array of tightly packed, hexagonal facets (lenses) and the array was similar in structure and appearance to insect eyes of today. There were ample amounts of calcite within the seawater so there was no shortage of calcite for the eye lens or exoskeleton of the trilobite. However, the eye lenses were made of an optically clear type of calcite now used in Nicol and Glan Thompson prisms to polarize light. Calcite is used for these prisms because it is a uniaxial, birefringent material. The trilobite's calcite eye lens has its optical axis oriented along the calcite c-axis, and as long as the field of view was narrow the calcite eye lens was not a problem. A further discussion of this point occurs in Section 4.

The index of refraction for calcite as for other transparent materials is a function of incoming light wavelength. For this paper the incoming light this wavelength is assumed to be 0.6 microns wavelength . Also, the extraordinary ray index of refraction axis is symbolized by $N_e$ whereas the ordinary ray axis is denoted by $N_o$. The index of refraction for index of refraction at angles for values in between these two extremes are calculated in Table 1,

Bragg [Bra 1924 a and b] studied calcite using x-ray analysis. Bragg's research is important because it defines the atomic structure of calcium and carbonate orientation within the crystal structure of calcite. From Bragg's research, the c-axis was determined. There are number of references for calcite optical data for example: [Cal 2012a and Cal 2012b].

Calcite is a negative birefringent crystal, which means that index of the extraordinary ray is larger than that for the ordinary ray index of refraction. At room temperature and 0.6 µ light, the ordinary index of refraction is 1.4860 and the extraordinary refraction is 1.6600 .

# 4. How the Index of Refraction for Calcite Varies with Light ray Propagation Angle

An occurrence the author experienced in an optical design employing a birefringent material is pertinent for the discussion here. The situation relates to a detector window made of sapphire, which is less birefringence then calcite. The detector vendor sealed the detector in a small can with a window made of sapphire. The vendor measured the precise focal plane for the detector surface through the sapphire window using near parallel light impinging perpendicularly to the window. However, the detector in operation was used in the image space where the incoming light to the detector was an F/1 cone (45°). The difference in propagation angle between the vendor's perpendicular incoming light and 45° the operational condition caused an intolerable blurring. This anecdote shows the attendant difficulty caused by birefringent materials when the light propagation angle is not along the optical axis.

The paper by Beyerie and McDermid [Bey 1998] shows the complexity involved for determining an electromagnetic theory for birefringent lenses. Other complex papers are by Lesso et al [Les 2000] and Eng and Leib [Eng 1969]. An interesting application for actually using the birefringent effect of calcite is the focus switching telescope by Kirby et al [Kir 2005]. In calcite, the index of refraction varies greatly, dependent upon the propagation light ray angle relative to the c-axis.. This point is illustrated In Table 1 . Table 1 was generated by applying the formula for the index of refraction variation from Nussbaum [Nus 1976 pp366]. This formula was derived on page 378 of [Nus 1976]. (For those who wish to see a more complete treatment of refraction in a uniaxial crystal see Born and Wolf , pp 811 to 818, [Bor 1959].

| Propagation Direction Relative to c-axis (degrees) | Index Of Refraction for Extraordinary Ray | Effective Focal Length (microns) | Back Focal Length (microns) | Angle From Paraxial Focal Point to Lens Edge, (U) ( radians) | Effectivr Focal Lenght at 0.000 minus calculated Effectivr Focal Lenght (microns) |
|---|---|---|---|---|---|
| 0.0000 | 1.6600 | 325.502 | 229.691 | 0.438 | 0.000 |
| 2.0 | 1.6600 | 325.719 | 229.850 | 0.424 | -0.219 |
| 4.0 | 1.6590 | 326.370 | 230.484 | 0.423 | -0.870 |
| 6,0 | 1.6578 | 327.456 | 231.543 | 0.421 | -1.956 |
| 8.0 | 1.6560 | 328.979 | 233.027 | 0.419 | -3.479 |
| 10.0 | 1.6538 | 330.941 | 234.940 | 0.418 | -4.441 |
| 12.0 | 1.6512 | 333.344 | 237.285 | 0.414 | -7.844 |
| 14.0 | 1.6480 | 336.194 | 240.065 | 0.410 | -10.694 |
| 16.0 | 1.6446 | 339.492 | 243.285 | 0.406 | -13.992 |
| 18.0 | 1.6407 | 343,244 | 246.949 | 0.402 | -17.744 |
| 20.0 | 1.6396 | 347.454 | 251.062 | 0.397 | -21.954 |
| 22.0 | --- | | | | |
| 24.0 | --- | | | | |
| 26.0 | --- | | | | |
| 28.0 | --- | | | | |
| 30.0 | 1 .6108 | 375.531 | 278.537 | 0.367 | -50.031 |
| 32.0 | -- | | | | |
| 34.0 | -- | | | | |
| 36.0 | -- | | | | |
| 38.0 | -- | | | | |
| 40.0 | 1.5810 | 415.550 | 317.811 | 0.332 | -90.050 |
| 42.0 | -- | | | | |
| 44.0 | -- | | | | |
| -- | -- | | | | |
| -- | -- | | | | |
| 90.0 | 1.4860 | 647.706 | 546.871 | 0.213 | -322.206 |

Table 1 – Summary Values for Calcite Lens Calculations

There are a number of things that can be gleaned from Table 1. First, notice the amount of the index of refraction variation as measured by the angle from the c-axis, in this case the extraordinary ray axis. At 0° in the index of refraction is 1.6600 and decreases as the propagation angle from the extraordinary ray axis increases. When 90° is reached, the index of refraction is 1.4800, the value for the ordinary ray axis. The important point is that when that the propagation angle within calcite is greater then about 10° there will be blurring of the image. Because the Effective Focal Length, F and Back Focal Length, BFL, vary with angle so that the size of light cone impinging about the c-axis causes a relative amount of blurring in the image. The values listed for U (in Table 1.) are important for determining the blur circle. Smith shows the many approximations relating to sine U and its relation to the various lens aberrations [Smi 1990 pp143-147 & 336-338].

The eye lens for the trilobite is modeled by a single, thick convex lens , (Figure 1). When the incoming rays are less than 5°, paraxial optics is valid. Horvath has written frequently on the topic of how the trilobite eye operates. In one of his papers he speculates about the values for the trilobite eye lens anatomy [Hov 1989]. This Horvath eye data are used here, but when a needed parameter is missing from the Horvath work, these data are estimated. The purpose here is to generate a generic evaluation for the trilobite eye lens. The calculation uses the lens and nomenclature is depicted in Figure 1. The lens calculations employ the paraxial method described by Nussbaum [Nus 1976, pp 13].

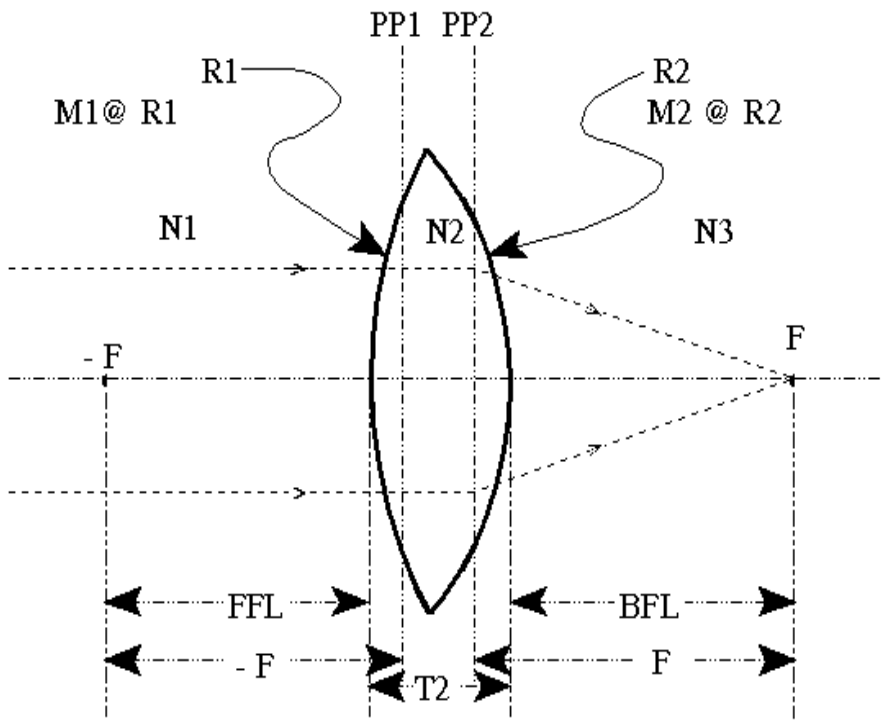

Figure 1 -Thick Lens Model for Eye Lens

The refracting power of the 1st surface is $K1=(N2-N1)/(R1)$ and likewise, for the 2nd surface $K2=(N3-N2)/R2$. Using K1 and K2 and the nomenclature from Figure 1 the Gaussian variables are :

$A = K1+K2-((K1*K2*T2)/N2)$
$B = 1-((K2*T2)/N2)$
$C = 1-((K1*T2)/N2)$

D= -T2/N2.

Using these Gaussian variables other lens parameters are calculated. For example, the Effective Focal Length, F=N3/(A) and the F/ number (F/#), which is F/AP where AP is the total lens aperture. Another important calculation is Back Focal Length, BFL which is N3*(C-1)/A . An interesting fact about the Gaussian variables is that=B*C-A*D=1 which can be used as a calculation check.

The parameters needed for the eye lens for the trilobite are the Horvath values and estimates by VLW.

> R1=135--- (from Horvath)
> R2=-135---(from Horvath)
> N1=1.33---(index of refraction for seawater)
> N2=1.66---(index of refraction for 0° from c-axis for calcite)
> N3=1.34---(index of refraction for ocular fluid)
> T1=1E6--- a great length distance in order to calculate focal length)
> T2=200--- (estimate by VLW since not given by Horvath)
> AP=276---(estimate by VLW from Horvath data).

On line 1 of Table 1 are the case for 0° from the optical axis or c-axis. The change of the eye lens focal length and other variables like, the index of refraction, vary according to the formula listed in [Nus 1976, pp 366] are listed as function of 2° per line. Using these variable values it is found that the focal length, F=325.502, the BFL=229.638 and F/#= 1.179 when the angle from the c-axis is 0°, the first line of Table 1.

The index of refraction listed in Table1 (column 2) is calculated in increments of 2° degrees up to 20°, and then listed as shown. Notice also that F, BFL and U are shown. The last column is very important because it shows how the Effective Focal Length varies with light propagation direction angle. For example, at 20° there is a difference of almost 22 µ difference in focal length from its value for 0°. Table 1 also shows that when the angle of propagation is 90° the index of refraction is the value for the ordinary ray. The values listed for U are important for considering the blur circle. Smith shows how sine U affects to approximate various lens aberrations [Smi 1990 pp143-147 & 336-338].

The most important thing to glean from Table 1 is that birefringence does cause a deleterious effect on the blur circle of the lens, which simulates the eye itself .

# 5. Box Jellyfish

The box jellyfish and the trilobite both are very ancient animals. However, there is a difference in their eye lens material. The box jellyfish has a crystallin eye lens for its eight camera-like eyes. (The box jellyfish has 24 eyes, eight camera-like eyes, eight pit eyes, and eight slit eyes.) The discussion here will concern the camera-like eyes, which are eye lenses of crystallin material. The box jellyfish of class Cubozoa is a primitive animal (see papers by Garm [Gar 2007, 2009, 2011], but it also has an unusual genome structure [Smi 2011].

Evolution of eyes are reviewed by Land and Nilsson [Lan 1992b] which discusses animal eyes in general. (Another review on the evolution of eyes is by Land and Fernald [Lan1992c,) An unusual trait of the box jellyfish is its eye placement. The eyes are mounted on a four-sided torrent (hence the name box), which allow the box jellyfish is see in four perpendicular directions simultaneously. Even though the box jellyfish's body is in soft and decays easily there has been a doubling of box jellyfish information in the last decade [Ben 2010]. An assumption is that present box jellyfish has changed little since the Cambrian era [Par 2003].

An important optical achievement was the optical measurement of the upper and lower camera-like lenses of the box jellyfish by Nilsson et al [Nil 2005]. The measurements were made by extracting the camera-like eyes from living animals and then measuring the lens parameters in the laboratory. Nilsson at al found that the image plane for both upper and lower eyes did not match the position of the focal plane. This point is further discussed by Hopkin [Hop 2005]. However, since the box jellyfish has no brain, and there is only a small amount image processing from its nerve net, the box jellyfish measures minimal views for the location of the shoreline. This being the case, the out of focus focal plane is of minor significance.

Relative to jellyfish, several points illustrate the thrust of this paper. First, the jellyfish camera-like eyes according to Garm, has a visual field-of-view of the order of 98° [Gar 2011]. Second, the jellyfish camera eyes are composed of a crystallin material not calcite. The author believes a lens material like crystallin was needed for vertebrates, in particular, rather than the trilobite calcite eye lens.

# 6. Discussion

The spider, although not a vertebrate, is an excellent example of an ancient animal with a crystallin eye lens. The spider has an eye lens material like that in future animals, a crystalline eye lens. (Some spiders have an acuity that approaches that of humans ( jumping spiders), and some spiders also have a large visual field of view.) Williams used the case of the spiders to demonstrate evolution [Wil 2010] and Wil [2011] by using a spider phenotype eye lens. Gehring [Geh 2010] contends that all eyes developed from one prototype eye, but prototype eye is still unknown. (See Langton [Lang 1997] for a discussion of phenotype and genotype) . Also, Gehring's hypothesis is still unproven.

A scientific theory like Darwin's evolution theory may be valid but not proven. An example is Einstein's General Theory of Relativity that is constantly being tested by researchers who try to find a condition where it is not true. Darwin himself has said if his theory[Dar 1859] is true, there are no cases which prove the falsity of the theory. Because evolution takes place over us such a long time periods, disproof is hard. One of the few ways evolution can be checked via computer simulation., this has been done by Williams [Wil 2010]. Strangely, about 40% of United Estates lay public think evolution cannot exist whereas about hundred percent of scientists think that evolution is both viable and true. Atmar [Atm 1997] lays out the conditions for evolution to be viable.

It would seem that there are two more important philosophical and religious questions to debate rather than debate the truth of evolution. These questions are are:
1. Why does life exist?
2. Why do those who have it, cling so violently to life.

A good starting point for these kinds of deliberations is Erwin Schrodinger's small book, "What is life?" [ Sch 1944].

# 7. Conclusion

Calcite for eye lenses is not found in modern animals. The animal kingdom, both for water and land animals, nowadays has a crystallin eye lens. This paper shows that a large field-of-view makes calcite eye lens very difficult to use.

REFERENCES

Note: the case for the title and the journal is determined by the journal's preference for case.

Bar 199 1    Barrow, J.D., Theories of Everything, Ballentine Books, p 19.

Bey 1998     Beyerie, Georg and McDermid, Sturart I, Ray-tracing formulas for refraction and internal reflection in uniaxial crystals, Applied Optics, Vol 37, No.34, pp7947-7953.

Box 2001     Amazing facts about the box jellyfish, ters to Nature, http://org.uk/corals-and-jellies/box-jellyfish.

Bor 1959     Born, Max and Wolf, Emil, Principles of optics, Cambridge, 952 pages.

Bra 1924a    The Influence of Atomic Arrangement on Refractive Index, Proc. R. Soc. A, 106, 346-368.

Bra 1924b    Bragg,W.L., The Refractive indices of Calcite and Aragonite, Proc.R. R. Soc. A, 105, 370-386.

Bru 2008     Bruckner, Andreas et al, Advanced artificial compound-eye imaging systems, Proc. SPIE,Vol.6887, 688709-0277-786X/08/$18.

Buc 1987     Buchsbaum, Ralph,et al, Animals Without Backbones, the University of Chicago press, Chicago, 572 pages.

Cal 2012a    Calcite, http://refractiveindex.info/?group=CRYSTAL.S&material=CaCo3

Cal 2012b    Calcite, Wickipedia, http://wickipedia.org/wiki/Calcite.

Cas 2009     Castagnoli, Guiseppe. Quantum computation and physical computation of biological information processing, arXiv:0912.548v1 [quant-ph] 30 Dec 2009.

Cha 1995     Chaitin, Gregory J., Randomness in arithmetic and the decline and fall of reductionism in pure mathematics, in Nature's Imagination, edited by John Cornwell, Oxford, p 41.

Clar 1979    CLARKSON , E.  N. K., Invertebrate Paleontology And Evolution, and George, Allen & Unwin, London, 323 pages.

Clar 1966    CLARKSON , E.  N. K., SCHIZOCHROAL EYES OF SOME SILURIAN ACASTID TRILOBITES, Paleontology, Vol. 9 Part pp 1-29.
9